\documentclass{article}

\begin{document}

{\bf CATEGORICAL GEOMETRY AND THE MATHEMATICAL FOUNDATIONS OF QUANTUM GRAVITY}

\bigskip

by Louis Crane, Mathematics Department Kansas State University

\bigskip

{\bf ABSTRACT:} {\it We consider two related approaches to
  quantizing general relativity which involve replacing point set
  topology with category theory as the foundation for the theory. The
  ideas of categorical topology are introduced in a way we hope is
  physicist friendly.}

\bigskip

{\bf I. INTRODUCTION}

\bigskip

The mathematical structure of a theory is a very abstract collection
of assumptions about the nature of the sphere of phenomena the theory
studies. Given the great cultural gap which has opened between
Mathematics and Physics, it is all too easy for these assumptions to
become unconscious.

General Relativity is a classical theory. Its mathematical foundation
is a smooth manifold with a pseudometric on it. This entails the
following assumptions:

\bigskip

1. Spacetime contains a continuously infinite set of pointlike events
   which is independent of the observer.

\bigskip

2. Arbitrarily small intervals and durations are well defined
   quantities. They are either simultaneously measurable or must be
   treated as existing in principle, even if unmeasurable.

\bigskip

3. At very short distances, special relativity becomes extremely
   accurate, because spacetime is nearly flat.

\bigskip

4. Physical effects from the infinite set of past events can all affect
  an event in their future, consequently they must all be integrated
  over.

\bigskip

The problem of the infinities in quantum general relativity is
intimately connected to the consequences of these assumptions.

In my experience, most relativists do not actually believe these
assumptions to be reasonable. Nevertheless, any attempt to quantize
relativity which begins with a metric on a three or four dimensional
manifold, a connection on a manifold, or strings moving in a
geometric background metric on a manifold, is in effect making them.

Philosophically, the concept of a continuum of points is an
idealization of the principles of classical Physics applied to the
spacetime location of events. Observations can localize events into
regions. Since classically all observations can be performed
simultaneously and with arbitrary accuracy, we can create infinite
sequences of contracting regions, which represent points in the limit.

Relativity and quantum mechanics both create obstacles to this
process. Determinations of position in spacetime cannot be arbitrarily
precise, nor can they be simultaneously well defined. 

Unfortunately the classical continuum is thousands of years old and is
very deeply rooted in our education. It tends to pass under the radar
screen.

I often suspect that quantum physicists are suspicious of Mathematics
because so much of it seems wrong to them. I think the solution is
more Mathematics rather than less.

One often hears from quantum field theorists that the continuum is the
limit of the lattice as the spacing parameter goes to 0. It is not
possible to obtain an uncountable infinite point set as a limit of finite
sets of vertices, but categorical approaches to topology do allow us
to make sense of that statement, in the sense that topoi of categories
of simplicial complexes are limits of them.

I have become convinced that the extraordinary difficulty of
quantizing gravity is precisely due to the omnipresence of the numerated
assumptions. For this reason, this paper will explore the problem of
finding the appropriate mathematical concept of spacetime in which a quantum theory
of GR could be constructed.

Now, although it is not well known among physicists of any stripe,
mathematicians have developed very sophisticated foundations both for
topology and the geometry of smooth manifolds, in which an underlying
point set is not required.

We will be interested in two related lines of development here;
higher category theory and topos theory. Over the last several years,
it has become clear that these mathematical approaches have a number
of close relationships with interesting new models for quantum
gravity, and also with foundational issues in quantum mechanics which
will have to be faced in QGR.

In this paper, I hope to introduce these ideas to the relativity
community. The most useful approach seems to be to begin with a
non-technical introduction to the mathematical structures involved,
followed by a survey of actual and potential applications to Physics.

\newpage

{\bf II. SOME MATHEMATICAL APPROACHES TO POINTLESS SPACE AND
  SPACETIME}
\bigskip

{\bf A. CATEGORIES IN QUANTUM PHYSICS; FEYNMANOLOGY}

\bigskip

Although categorical language is not explicitly familiar in Physics,
quantum field theory is in fact dominated by the theory of tensor
categories under a different name. A category is a mathematical
structure with objects and maps between them called morphisms [1]. A
tensor category has the further structure of a product which allows us
to combine two objects into a new one.

If we write out a general morphism in a tensor category, we get arrows
starting from sets of several objects and ending in different sets of
several objects, where we think of maps into and out of the tensor
product of objects as maps into and out of their combination. When we
compose these, we get exactly Feynman graphs.

The objects in the category are the particles (or more concretely
their internal hilbert spaces), and the vertices are the tensor
morphisms.

The resemblance is not accidental. The kronecker product which tells
us how to combine the hilbert spaces of subsystems is just what
mathematicians call the tensor product. 

The representations of a lie group form a tensor category, in which
the morphisms are the maps which intertwine the group action. This
is equivalent to the prescription in feynmanology that we include all
vertices not excluded by the symmetries of the theory.

The physicist reader can substitute for the idea of a categorical
space the idea that the spacetime is actually a superposition of
Feynman graphs, which we can think of as a vacuum fluctuation. 

The feynmanological point of view has been developed for the BC model
under the name of group field theory [2]. The 4-simplices of the
triangulation are treated as vertices in this point of view, and the
3-simplices as particles. 

The categorical language is much more developed, and connected to more
mathematical examples. I hope I will be forgiven for staying with it.

\bigskip

{\bf B. GROTHENDIECK SITES AND TOPOI}

\bigskip

The mathematical ideas we shall consider trace back to the work of
Alexander Grothendieck, perhaps the deepest mathematical thinker who
ever lived. Much of his work was only appreciated after several
decades, his deepest ideas are still not fully understood.

Grothendieck made the observation that the open sets of a topological
space could be considered as the objects of a category, with a
morphism between two objects if the first was contained in the second.
He called this the site of the space. This was motivated by the
observation that presheaves over the space are the same as functors
from the site to the category of whatever type of fiber the sheaf is
supposed to have. Since the constructions of topology and geometry can
be reformulated in terms of presheaves, (a bundle, for example can be
replaced with the presheaf of its local sections), this opened the way
to a far ranging generalization of topology and geometry, in which
general categories play the role of spaces.

Grothendieck also realized that rather than the site itself, the
central object of study was the category of presheaves over it, (or
functors into the category of sets), which
he called its topos [3].

Topoi also have an axiomatic definition, which amounts to the idea
that they are a category in which all the normal constructions done on
sets have analogs. It was then proven that every abstract topos is the
topos of some site [3,4]. 

For this reason, the objects in a topos can be thought of either as
abstract 
sets, or
variable or relative sets. 

One of the interesting aspects of topos theory is that the objects in
a topos can inherit structure from the objects in the category which
is its site.

An important example is synthetic differential geometry [5], the study
of the topos over the site category of smooth rings, or ``analytic
spaces'' (there are several variants).

Objects in this topos inherit a notion of differential and integral
calculus. The object in this category which corresponds to the real
numbers has infinitesimal elements. It is much more convenient to
treat infinitesimals in a setting where not everything is determined
by sets of elements. The result is that the calculus techniques of
physicists which mathematicians are forever criticizing suddenly
become rigorous.

A topos is a more subtle replacement of the notion of space than a
category. It is a category of maps between categories, so it has the
character of a relative space. In this paper, we are exploring the
possibility that the relativity of objects in a topos could be a model
for the relativity of the state of a system to the observer.

\bigskip

{\bf C. HIGHER CATEGORIES AS SPACES}

\bigskip
The idea that topology and geometry are really about regions and maps
between them rather than sets of points, has been a subtle but
widespread influence in Mathematics. 

A mathematical object with many objects and maps between them is a
category [1]. There are many approaches to regarding a category as a kind
of space.

Mathematicians have extended the idea of a category to an
n-category. A 2-category has objects, maps and maps between maps,
known of as homotopies or 2-morphisms. An n-category has 1,2..n
morphisms [6]. 

The simplest situation in which a higher category can be thought of 
as a
kind of space is the case of a simplicial complex.

A simplicial complex is a set of points, intervals, triangles
tetrahedra etc referred to as n-simplices, where n is the
dimension. The faces of the n-simplices are identified with n-1
simplices, thus giving a discrete set of gluing rules. Faces are
defined combinatorially as subsets of vertices. The whole structure is
given by discrete combinatorial data.

A simplicial complex is thus a discrete combinatorial object. It does
not contain a sets of internal points. These can be added to form the
geometric realization of a simplicial complex, but that is usually not
done. 

Because the vertices of a simplex are ordered, which fixes an
orientation on each of its faces of all dimensions, it is natural to
represent it as a higher category. The vertices are the objects, the
edges are 1-morphisms, the triangles 2-morphisms etc.

For many purposes, simplicial complexes are just as good as
topological spaces or manifolds. Physicists who like to do Physics on
a lattice can generalize to curved spacetime by working on a
simplicial complex.

There is also a notion of the topology of a simplicial complex
including cohomology and homotopy theory. A celebrated theorem states
that the categories of homotopy types of simplicial complexes and of
topological spaces are equivalent [7].

A naive first approach to quantum spacetime would say that at the
Planck scale spacetime is described by a simplicial complex, rather
than a continuum. This point of view would nicely accommodate the
state sum models for quantum gravity, and the categorical language
would allow a very elegant formulation of them, as we shall discuss
below. The richness of the connections between category theory and
topology allows for more sophisticated versions of this, in which
simplicial complexes appear relationally, ie. the information flowing
between two regions forms a simplicial complex. We will discuss
physical approaches to this below.

Another way to relate categories to simplicial complexes is the
construction of the nerve of a category, which is a simplicial complex
which expresses the structure of the category. The nerve is
constructed by assigning an n-simplex to each chain of n+1 composible
morphisms in the category. The n-1 faces are each given by composing
one successive pair of morphisms to form an n-chain.

The simplicial complex so formed is a generalization of the
classifying space of a group. A group is a category with one object
and all morphisms invertible.

There are also constructions which associate a category to a cellular
or cubical complex.

The various descriptions of spaces by categories also extend to
descriptions of maps between spaces as functors between categories.

Since the setting of a Yang-Mills or Kaluza-Klein theory is a
projection map
between manifolds, these have categorical generalisations which
include more possibilities than the manifold versions.

One very interesting aspect of topos theory is the change in the
status of points. A topos does not have an absolute set of points;
rather, any topos can have points in any other topos. This was
originally discovered by Grothendieck in algebraic geometry [3], where the
topoi are called schemes. We shall discuss physical implications of
this below.
\bigskip

{\bf D. STACKS AND COSMOI}

\bigskip
As we shall see in the next section, both higher categorical and topos
theoretical notions of space have strong connections to ideas in
quantum gravity. For various reasons, it seems desirable to form a
fusion of the two; that is, to form relational versions of higher
categories.

Interestingly, this was the goal of the final work of Grothendieck
on stacks, which he did not complete. Much of this has been worked out
more recently by other authors [8].

The maps between two categories form a category, not merely a
set. This is because of the existence of natural transformations
between the functors. Similarly the morphisms between two 2-categories
form a 2-category etc. The analog of sheaves over sites for
2-categories are called stacks. Much as the case of sheaves, these are
equivalent to 2-functors. Incidently the word Grothendieck chose
for a stack in French is champs, the same as the French word for a
physical field.

One can also investigate the 2-categorical analog for a topos, which
is a 2-category with an analogous structure to the 2-category of all
``small'' categories. This has been defined under the name of a
cosmos [9]. 

An interesting class of examples of stacks are the gerbes [10], which have
attracted interest in string theory and 2-Yang-Mills theory [11]. Theories
with gerbe excitations would generalize naturally into a 2-categoric
background spacetime.

\bigskip

{\bf III. PHYSICS IN CATEGORICAL SPACETIME}

\bigskip
The ideal foundation for a quantum theory of gravity would begin 
with a
description of a quantum mechanical measurement of some part of 
the geometry of
some region; proceed to an analysis of the commutation relations
between different observations, and then hypothesize a mathematical
structure for spacetime which would contain these relations and give
general relativity in a classical limit.

We do not know how to do this at present. However, we do have a number
of approaches in which categorical ideas about spacetime fit with
aspects of geometry and and quantum theory in interesting ways. We
shall present these, and close with some ideas about how to achieve a
synthesis.

\bigskip

{\bf A. THE BC CATEGORICAL STATE SUM MODEL}

\bigskip

The development of the Barrett-Crane model for quantum general
relativity [12, 13] begins by substituting a simplicial complex for a
manifold. It is possible to adopt the point of view that this is
merely a discrete approximation to an underlying continuous geometry
located on a triangulation of the manifold. That was never my
motivation.
Rather considerations of the Planck scale cutoff and the limitations
of information transfer in general relativity suggested that discrete
geometry was more fundamental.

In any event, the problem of quantizing the geometry on a simplicial
complex has proved to be much more tractible than the continuum
version.

The bivectors assigned by the geometry to the triangles of the complex
can be identified with vectors in the dual of the lorentz algebra, and hence have
a very well understood quantization using the Kostant-Kirillov
approach [14]. The quantum theory reduces to a careful combination of the
unitary representations of the lorentz algebra due to Gelfand [15,16],
and 
of intertwining operators between them.

We tensor together the representations corresponding to the
 assignments
 of area
variables to the faces, then take the direct sum over all
labellings. The resultant expression is what we call a categorical
state sum.

The expression obtained for the state sum on any finite simplicial
complex has been shown to be finite [17].

In addition, the mathematical form of the state sum is very elegant
from the categorical point of view. If we think of the simplicial
complex as a higher category, and the representations of the lorentz
group as objects in a tensor category (which is really a type of
2-category), then the state sum is a sum over the functors between
them. 

The BC model is expressed as the category of functors between a
spacetime category and a field category. The field category being a
suitable subcategory of the unitary representations of the lorentz
algebra.
This suggests a general
procedure 
for connecting more sophisticated categorical approaches to spacetime
to
quantum gravity. Namely, we could examine the category of functors
from whatever 
version of spacetime category we are studying to the representation
category of the lorentz algebra in order to put in the geometric
variables. 

It is not necessary for the simplicial complex on which we define the
BC model to be equivalent to a triangulation of a manifold. A 4D
simplicial complex in general has the topology of a manifold with
conical singularities. There has been some work interpreting the
behavior of the model near a singular point as a particle, with
interesting results [18, 19]. The singularities conic over genus 1 surfaces
reproduce, at least in a crude first approximation, the bosonic sector
of the standard model, while the higher genus singularities decouple
at low energy, with interesting early universe implications.
The possibility of investigating singular points would not arise in any theory
formulated on a manifold.

The BC model has not yet gained general acceptance as a candidate for
quantum general relativity. The fundamental problem is the failure of
attempts to find its classical limit. 

I want to argue that the work done to date on the classical limit of
the model, my own included, has been based on a misconception.

A categorical state sum model is not a path integral, although it
resembles one in many aspects. Rather the geometry of each simplex has
been quantized separately, and the whole model represented on a
constrained tensor product of the local hilbert spaces.

For this reason the terms in the CSS are not classical histories, but
rather quantum states. It is not really surprizing, then, that the
geometric variables on them do not have simultaneous sharp values, or
that they can contain singular configurations. Attempting to interpret
them as classical is analogous to confusing the zitterbewegung of the
electron with a classical trajectory.

In order to construct the classical limit of the BC model, it is
necessary to study the problem of the emergence of a classical world
in a quantum system. Fortunately, there has been great progress on
this in recent years in the field variously known as consistent
histories or decoherence.

The decoherent or consistent histories program has recently been interpreted as
indicating that quantum measurements should be considered as occurring
in a topos.

In the next sections, we shall briefly review the ideas of consistent
histories and decoherence, and explain how they lead to topos theory.
Then we shall discuss how to apply these ideas to the BC model.
\bigskip

{\bf B. DECOHERENT HISTORIES AND TOPOI}

\bigskip

The consistent histories/decoherence approach to the
interpretation of quantum mechanics is concerned with the problem of
how classical behavior emerges in a suitable approximation in a
quantum system [20].

We have to begin by coarse graining the system to be studied by
decomposing its hilbert space into a sum of subspaces described as the
images of orthogonal projections. A history is a sequence of members
of the set of projections at a sequence of times.

Next we need to define the decoherence functional D. It is the trace of 
the product of the first series of projections time reversed, the
density matrix of the original state of the system, and the product of
the first series of projections.

\bigskip

$D(H_1,H_2)=tr({H_1}^* \rho H_2)$

\bigskip

Classical behavior occurs if the decoherence functional is
concentrated on the diagonal, more precisely if there is a small
decoherence parameter $\eta$ such that

\bigskip

$D(H_1, H_2)= o( \eta)$ if $H_1 \neq H_2$.

\bigskip

This implies that states described by histories from the chosen set do
not interfere significantly. This implies classical behavior.

The next property of D to prove is that it concentrates near histories
which correspond to solutions of the equations of motion. This is a way
of affirming the correspondence principle for the system.

Since consistency is not perfect, we must think of the classical limit
as appearing in the limit of coarse grainings.

Decoherence, the second half of the program, is an extremely robust
mechanism causing histories to become consistent. When the variables
correspond to typical macroscopic quantities decoherence occurs
extremely quickly.

The central observation of the decoherence program is that classical
systems can never be effectively decoupled from their environment.

For instance, a piston in a cylinder containing a very dilute gas
might experience a negligible force. Nevertheless, the constant
collisions with gas molecules would cause the phase of the piston,
treated as a quantum system, to vary randomly and uncontrollably. 

Since it is not possible to measure the phases of all the molecules,
the determinations an observer could make about the position of the
piston would be modelled by projection operators whose images include
an ensemble of piston states with random phases, coupled to gas
molecule states.

This effect causes pistons (or any macroscopic body)  
to have diagonal decoherence functionals to
a high degree of accuracy, and hence to behave classically.

The definition of a classical system as one which cannot be
disentangled is a very useful one. It has enabled experiments to be
designed which study systems which are intermediate between classical
and quantum behavior [20].

When we observe a system, it is not possible to say exactly what set
of consistent histories we are using. It is more natural to think that
we are operating in a net of sets of consistent histories
simultaneously. 

We then expect that the result of an observation will be consistent if
we pass from one set of consistent histories to a coarse graining of
it.

The idea has been studied that this means that the results of
experiments should be thought of as taking values in a topos [21]. The
category whose objects are sets of consistent histories and whose
morphisms are coarse grainings can be thought of as a site, and the
results of experiments take place in presheaves over it.

In my view, the implications of this idea should be studied for
physical geometry. Does it mean, for example, that the physical real
numbers contain infinitesimals?

\bigskip

{\bf C. APPLICATION OF DECOHERENT HISTORIES TO THE BC MODEL}

\bigskip

This section is work in progress.

We would like to explore classical histories in the BC model. The goal
of this is to show that consistent histories exist for the model which
closely approximate the geometry of pseudoriemannian manifolds, and
that the decoherence functional concentrates around solutions of
Einstein's equation.

The natural choice for macroscopic variables in the BC model would be
the overall geometry of regions composing a number of simplices in the
underlying complex of the model. It is easier to choose the regions
themselves to be simplices which we call large to distinguish them
from the fundamental simplices of which they are composed.

The program for showing that the geometric data on the internal small
simplices decoheres the overall geometry of the large ones involves
two steps.

In the first, we use microlocal analysis to construct a basis of
states in which all the geometrical variables of the large simplices
are simultaneously sharp to a small inaccuracy. These would combine to
give a set of projection operators whose images correspond to
pseudoriemannian geometries on the complex, now thought of as a
triangulated manifold.

This problem is mathematically similar to finding a wavepacket for
a particle. The symplectic space for the tetrahedron turns out to be
equivalent to the symplectic structure on the space of euclidean
quadrilaterals in the euclidean signature case, and to have an
interesting hyperkahler structure in the case of the lorentzian
signature. This allows us to use powerful mathematical
simplifications, which make me believe the problem is quite solvable.

The second step would be to show the decoherence functional which
arises from averaging over the small variables causes the large
variables to decohere, and that the decoherence functional
concentrates around solutions of Einstein's equation.

This is quite analogous to known results for material systems such as
the piston. 

The existence of a Brownian motion approximation for the internal
variables makes me hopeful that this will work out, similarly to the
case of the piston, where an ideal gas approximation is the key to the
calculation. 

A more challenging problem would be to work out the topos theoretic
interpretation of the decoherence program in the case of the BC model.

The site of this topos would be the category whose objects are the
``large'' triangulations, and whose morphisms are coarse grainings.
 
One could then apply the ideas about modelling quantum observation in
a topos described above to the BC model. This would amount to the
construction of a 2-stack, since the BC model itself is 2-categorical.

This would give us a setting to ask the question: ``what does one
region in a spacetime, treated as classical, observe of the geometry of 
another part?''

This problem was suggested to me by Chris Isham.

\bigskip

{\bf D. CAUSAL SITES}
\bigskip

As we explained above, the site of a topological space X is a category
whose objects are the open sets of X and whose morphisms are
inclusions. The whole construction of a site rests on the relationship
of inclusion, which is a partial order on the set of open
subsets. This change of starting point has proven enormously
productive in Mathematics.

In Physics up to this point, the topological foundations for spacetime
have been taken over without alteration from the topological
foundation of space. In general relativity, a spacetime is
distinguished from a four dimensional space only by the signature of
its metric. 

Categorical concepts of topology are richer and more flexible than
point sets, however, and allow specifically spacetime structures to
become part of the topological foundation of the subject.

In particular, regions in spacetime, in addition to the partial order
relation of inclusion, have the partial order relation of causal
priority, defined when every part of one region can observe every part
of the other. 

The combination of these two relations satisfy some interesting
algebraic rules. These amount to saying that the compact regions
of a causal spacetime are naturally the objects of a two category, in
much the same way that open sets form a site.

This suggests the possibility of defining a spacetime directly as a
higher categorical object in which topology and causality are
unified. A topodynamics to join geometrodynamics.

Recently, Dan Christensen and I implemented this proposal by giving a
definition of causal sites and making an investigation of their
structure [22].

We began by axiomatizing the properties of inclusion and causal
order on compact regions of a strongly causal spacetime, then looked
for more general examples not directly related to underlying point sets.

The structure which results is interesting in a number of ways. There
is a natural 2-categorical formulation of causal sites. Objects are
regions, 1-morphisms are causal chains, defined as sequences of
regions each of which is causally prior to the next, and 2-morphisms
are inclusions of causal chains, rather technically defined. 

We think of causal chains as idealisations of observations, in which
information can be retransmitted.

We discovered several interesting families of examples. One family was
constructed by including a cutoff minimum spacetime scale. These
examples have the
interesting property that the set of causal chains between any two
regions has a maximal length. This length can be interpreted as the
duration of a timelike curve, and can very closely approximate the
durations in a classical causal spacetime.

Since the pseudometric of a spacetime can be recovered from its
timelike durations, the 2-categorical structure of a causal site can
contain not only the topology of a spacetime, but also its geometry.

We also discovered that any two causally related regions have a
relational tangent space, which describes the flow of information
between them. This space has the structure of a simplicial complex, as
opposed to a causal site itself, which has a bisimplicial structure
because of the two relations on it. In category theoretic terms, the
spacetime is a 2-category, but relationally it is a category.

An interesting feature of causal sites is that regions have relational
points, i.e. regions which appear to another region to be
indivisible, but perhaps are not absolutely so.

We hope that this feature may make causal sites useful in modelling
the theory of observation in general relativity, in which only a
finite amount of information can flow from one region to another [23], so
that an infinite point set is not observably distinguished.

We also think it an interesting echo of the relational nature of
points in topos theory.

If infinite point sets cannot be observed, than according to
Einstein's principle, they should not appear in the theory. Causal
sites are one possible way to implement this.

\newpage

{\bf E. THE 2-STACK OF QUANTUM GRAVITY? FURTHER DIRECTIONS}

\bigskip

At this point, we have outlined two approaches to categorical
spacetime, which include geometric information corresponding to the
metric structure in general relativity in two different ways.

In the Barrett-Crane model, the data which expresses the geometry is
directly quantum in nature. The geometric variables are given by
assigning unitary representations of the lorentz algebra to the
2-faces or triangles of a simplicial complex. These are hilbert spaces
on which operators corresponding to elements of the lorentz algebra
act, thus directly quantizing the degrees of freedom of the bivector,
or directed area element, which would appear on the 2-face if it had a
classical geometry, inherited from an embedding into Minkowski space.

This model also has a natural functorial expression, as we mentioned above.

On the other hand, in the causal sites picture spacetime is
represented as a family of regions, with two related partial orders on
them. Mathematically this can be expressed by regarding the regions as
the vertices of a bisimplicial set. Bisimplicial sets are one
mathematical approach to 2-categories [24]. This is also an expression of
the topological structure of the spacetime, although a more subtle one
than a simplicial complex, which could arise from a triangulation of a
manifold.

In some interesting examples, the {\bf classical} geometry of a spacetime 
is naturally included in this bisimplicial complex, measured by the
lengths of maximal causal chains. The approximation of the geometry by
a causal set [25] can not be as precise, since a causal site has minimal
regions which can be adapted to the direction of a path.

Now how could these two picture be synthesized?

One element which has not been included so far in the structure of a
causal site is local symmetry. It is clear that this would have to
appear in a fully satisfactory development of the theory, since the
local symmetries of spacetime are so physically important.

Including local symmetry in the structure of a causal site seems a
natural direction to study in linking the causal sites picture to the
BC model, since the geometrical variables of the BC model are
representations of the lorentz group. 

The fundamental variables of a causal site have a yes/no form: region
A either is or is not in the causal past of region B. 

We could attempt to quantize a causal site by replacing the definite
causal relations by causality operators.  We can now define a 2-dimensional
hilbert space H(A,B) for each pair of regions with a basis representing
the yes and no answers to the causal relatedness question. This
corresponds to a gravitational experiment in which an observer at B
sees or fails to see an event at A. The totality of such experiments
should define a quantum geometry on the site in the cases discussed
above with bounds on chain length, since the metric can be
effectively reconstructed from the classical answers. 

In the presence of an action of local symmetry on the regions of the
site, the tensor product of the spaces H(A,B) would decompose into
representations of the local symmetry group.

If this led to the reappearance of the BC model on the relative
tangent space between two regions in a site, it would create a setting
in which the idea of the BC model as describing the geometry of one
region as observed by another could be realized.

The physical thought is that since only a finite amount of information
can pass from A to B in general relativity, the set of vertex points
in a relative BC model could include all the topology of A which B
could detect.

The idea of construcing a topos version of the BC model using
decoherent histories also points to a BC model which varies depending
on which classical observer the model is observed by.

Both of these ideas (neither implementsd yet, and neither easy) seem
to hint at a simultaneously higher categorical and topos theoretical
description of quantum spacetime which would fulfill the physical idea
of a relational spacetime.

Perhaps there is an as yet unguessed construction of a 2-stack which
will provide a synthesis of these ideas. Einsteins relational ideas
may find their final form in the mathematical ideas of Grothendieck.

\bigskip

{\bf ACKNOWLEDGMENTS:} The idea of topos theory arising in quantum
theory in general and quantum gravity in particular is something I
learned fron Chris Isham. Much of the higher category theory in this
paper was strongly influenced by working with Dan Christensen durring
my visit to the University of Western Ontario. I learned about
Grothendieck's work durring my visit to Montpellier where I was
invited by Philippe Roche. I benefited from conversations about topos
theory with Carlos Contou-Carrere while I was there. I also had many
interesting conversations with Marni Sheppeard at both places.
The BC model, of course, is joint work with John Barrett.

\bigskip

{\bf BIBLIOGRAPHY}

\bigskip
1. S. Maclane, Categories for the working mathematician Springer
   Verlag, N.Y. 1971.

\bigskip

\bigskip
2. M. P. Reisenberger and C. Rovelli, Space time as feynman diagrams,
   the connection formulation, CQG 18 (2001) 121-140.

\bigskip

\bigskip
3. M. Artin, Theorie des topos et cohomologie etale des schemas,
   Springer, Berlin, N.Y. 1972.

\bigskip

\bigskip
4. S. Maclane and I. Moerdijk, Sheaves in geometry and logic, a first
   introduction to topos theory, Springer N. Y. 1992

\bigskip

\bigskip
5. A. Kock Synthetic differential geometry, Cambridge University
   Press, 1981.

\bigskip

\bigskip
6. J. Benabou, Introduction to bicategories in Reports of the
   midwest category theory seminar  LNM 47 Springer, 1967 p1-77.

\bigskip

\bigskip
7. P. Goerss and R. Jardine, Simplicial homotopy theory, Birkhauser, 1999

\bigskip

\bigskip

8. L. Breem, On the classification of 2-Gerbes and 2-Stacks, Societe
   Mathematique de France, Providence RI, 1994

\bigskip

\bigskip
9. R. Street, Cosmoi of internal categories, AMS Transactions, 1980

\bigskip

\bigskip
10. R. Picken A cohomological description of abelian bundles and
    gerbes proceedings XXth workshop on geometric methods in physics,
    Bielowieza, July 1-7 2001.

\bigskip

\bigskip
11. J. Baez and J. Dolan, Higher Yang-Mills theory, hep-th 0206130

\bigskip

\bigskip
12. J. Barrett and L. Crane, Relativistic spin networks and quantum
    gravity, J. Math. Phys. 39 (1998) 3296-3302

\bigskip

\bigskip
13. J. Barrett and L. Crane, A lorentzian signature model for quantum
    general relativity, CQG 17(2000) 3101-3118

\bigskip

\bigskip

14. J. C. Baez and J. Barrett, The quantum tetrahedron in 3 and 4
    dimensions, Adv. Theor. Math. Phys. 3, 815-850, 1999

\bigskip

\bigskip
15. G. Gelfand et. al. Generalized functions vol. 5 Integral geometry
    and representation theory, Academic Press 1966

\bigskip

\bigskip

16. I. M. Gelfand and M. A. Naimark, Unitary representations of the
    proper lorentz group, Izv. Akad. Nauk. SSSR  11 411 (1947)

\bigskip

\bigskip

17. L. Crane, A. Perez and C. Rovelli, A finiteness proof for the
    lorentzian state sum model for quantum general relativity,
    Phys. Rev. Lett. 87(2001) 181301

\bigskip

\bigskip
18. L. Crane A new approach to the geometrization of matter, gr-qc
    0110060.

\bigskip

\bigskip

19. S. Alexander, L. Crane, M. D. Sheppeard, The geometrization of
    matter proposal in the Barrett Crane model and resolution of
    cosmological problems gr-qc 0306079.

\bigskip

\bigskip
20. R. Omnes, Understanding quantum mechanics, Princeton University
    Press, 1999.

\bigskip

\bigskip
21. C. J. Isham, J. Butterfield, Some possible roles for topos theory
    in quantum theory and quantum gravity, Found. Phys. 30 (2000) 1707-1735.

\bigskip

\bigskip
22. J. D. Christensen and L. Crane, Causal sites and quantum gravity,
    to appear JMP.

\bigskip

\bigskip
23. M. A. Perlath, R. M. Wald, Comment on entropy bounds and the
    generalized second law, Phys. Rev. D. 60 (1999) 104009.

\bigskip

24. Z. Tamsamani Sur les notions de $ \infty$ categorie et $
    \infty$ groupoide non-stricte via des ensemble
    multi-simpliciaux, alg-geom 9512006

\bigskip

25. R. Sorkin, Causal sets, Discrete gravity, gr-qc 0309009

\end{document}